\documentclass[pra,aps,twocolumn,10pt,superscriptaddress,longbibliography,floatfix]{revtex4-2}

\usepackage[dvipsnames]{xcolor}
\usepackage{graphicx}
\usepackage{amsmath,amssymb}
\usepackage{bm}
\usepackage{braket}
\usepackage[colorlinks=true,linkcolor=blue!60!black,citecolor=blue!60!black,urlcolor=blue!60!black]{hyperref}

\providecommand{\selectlanguage}[1]{}
\newcommand{\Cbar}{\overline{\mathcal{C}}}
\newcommand{\Nbar}{\overline{N}}
\newcommand{\ii}{\mathrm{i}}
\newcommand{\dd}{\mathrm{d}}

\begin{document}

\title{Symmetry-selected entanglement death at Floquet resonances of parametrically coupled qubits}

\author{Kamran~Akbari}
\email[]{kamran.akbari@queensu.ca}
\affiliation{Department of Physics, Engineering Physics and Astronomy, Queen's University, Kingston ON K7L 3N6, Canada}
\author{Stephen~Hughes}
\email[]{shughes@queensu.ca}
\affiliation{Department of Physics, Engineering Physics and Astronomy, Queen's University, Kingston ON K7L 3N6, Canada}
\date{\today}

\begin{abstract}
Periodic modulation of a qubit--qubit coupling is a standard way to engineer quantum interactions, but a strong modulation also reorganizes the transitions through which the qubits dissipate into a comb of Floquet channels. We study two
identical, strongly coupled qubits whose transverse coupling is modulated anharmonically, as by a nonlinear coupler, with each qubit damped by its own reservoir. The modulation raises the ground-state concurrence of the cycle-averaged Hamiltonian from 0.45 to above 0.8, yet the steady-state concurrence stays close to its undriven value except at Floquet resonances
selected by symmetry. There it undergoes entanglement death, vanishing at all times in the drive cycle over a finite range of modulation depths. Parity and exchange symmetry split the Floquet spectrum into an even sector, whose branches
anticross at multiphoton resonances, and an exactly solvable odd sector, whose branches cross. Every dissipative channel connects the two sectors, so only the
even anticrossings cause entanglement death, provided their gap exceeds the damping rate, whereas the nearby odd crossings leave the concurrence unchanged.
Comparing time-independent and Floquet master equations, with and without the secular approximation, we show that the basis in which the dissipator is built
changes the entanglement by up to 0.2 for an Ohmic bath and 0.5 for a flat one, whereas the secular approximation matters only near anticrossings with gaps comparable to the damping. The basis matters most when the bath has spectral weight at low frequency, where the drive moves dissipative channels through zero
frequency. A transverse field that breaks parity lifts the selection rule.
\end{abstract}

\maketitle

\section{Introduction}
\label{sec:intro}

\begin{figure*}[t]
\centering
\includegraphics[width=\textwidth]{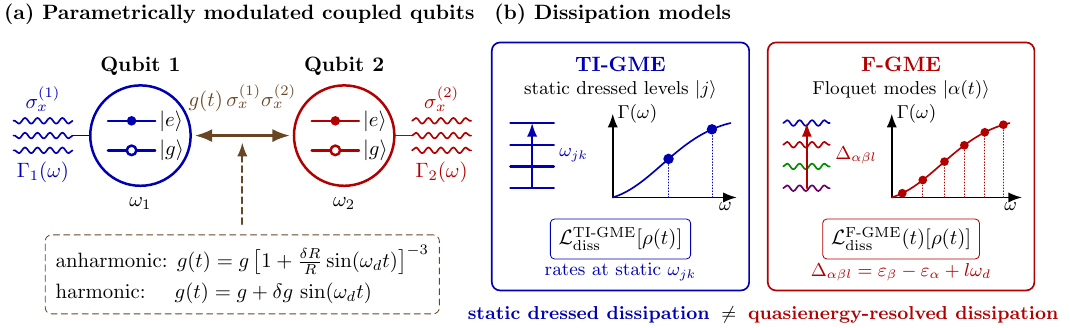}
\caption{Model schematic.
(a)~Two qubits with transverse coupling $g(t)\sigma_x^{(1)}\sigma_x^{(2)}$,
periodically modulated at frequency $\omega_d$, each coupled to its own reservoir
$\Gamma_\Lambda(\omega)$ through $\sigma_x^{(\Lambda)}$. The main results use the
anharmonic modulation; the harmonic one is used in Sec.~\ref{sec:harmonic}.
(b)~Structure of the dissipators. The TI-GME samples the bath at the static
dressed transition frequencies $\omega_{jk}$, whereas the F-GME samples it at the
quasienergy channels
$\Delta_{\alpha\beta l}=\varepsilon_\beta-\varepsilon_\alpha+l\omega_d$, which can
lie close to zero frequency.}
\label{fig:model}
\end{figure*}

Periodic driving is a versatile tool for engineering effective Hamiltonians,
including interactions and entanglement that are absent in the static
system~\cite{Bukov_Universal_2015,Eckardt_Atomic_2017,Maskara_Entanglement_2021,
Petiziol_ThreeBody_2021,Koyluoglu_Floquet_2025,Oka_Floquet_2019,Silveri_Quantum_2017}. In superconducting circuits,
modulating a tunable coupler activates resonant interactions between otherwise
detuned qubits, and this is the basis of parametric two-qubit
gates~\cite{Bertet_Parametric_2006,Niskanen2007,Yan_Tunable_2018,
Han_ParametricISWAP_2020,Mitchell_Hardware-Efficient_2021,McKay_Universal_2016,
Reagor_Demonstration_2018,Caldwell_Parametrically_2018,Sung_Realization_2021,
Sete_Parametric_2021}.
Any real device, however, is an open quantum system, coupled to its environment
through dissipation and decoherence.
 Dissipation usually degrades entanglement, but it
can also be a resource. Engineered reservoirs can prepare and stabilize
entangled steady states~\cite{Poyatos_Quantum_1996,Kraus_Preparation_2008,
Verstraete_Quantum_2009,Krauter_Entanglement_2011,Lin_Dissipative_2013,
Shankar_Autonomously_2013,Reiter_SteadyState_2013,Harrington_Engineered_2022},
so the steady state of a driven, damped system is a natural target for state
engineering.

In an open system, the drive that reshapes the coherent dynamics also reorganizes
the transition frequencies and matrix elements through which the environment
acts. The relevant transitions are then between Floquet states, and their
frequencies are quasienergy differences shifted by multiples of the drive
frequency~\cite{Grifoni_Driven_1998,Kohler_FloquetMarkov_1997,Kohler_Driven_2005,
Hausinger_Dissipative_2010,Mori_FloquetOpenSystems_2023,shirai_floquetgibbs_2018,
Hone_Statistical_2009,Gasparinetti_Environment_2013}.
When a coupling is a sizable fraction of the bare transition frequencies,
typically above about ten percent, the system is in the ultrastrong-coupling
regime. The term is usually applied to light--matter systems, such as a qubit in a
cavity~\cite{FriskKockum_Ultrastrong_2019,FornDiaz_Ultrastrong_2019,
Niemczyk_Circuit_2010,Yoshihara2017}, but the same
regime is reached when two qubits are coupled directly, which is the case
considered here. Both subsystems are then matter-like, and the counter-rotating
terms of their coupling can no longer be neglected. The ground state contains
virtual excitations of both qubits, and the dissipator must be built from
interaction-dressed rather than bare states~\cite{Ciuti_InputOutput_2006,
Beaudoin_Dissipation_2011,Settineri_Dissipation_2018,Salmon_Gauge-independent_2022,
Akbari_Generalized_2023}. The system is then dressed twice, internally by the
interaction and externally by the
drive~\cite{Akbari_Floquet_2025,Akbari_Floquet-Liouville_2026,Akbari_Floquet_2026}.

This raises a simple question. Does an effective Hamiltonian with a more
entangled ground state produce a more entangled steady state? 
A common practical
approach keeps the static dressed dissipator and adds the drive only to the
coherent part of the dynamics. 
Pacino \textit{et al.}~\cite{Pacino_Foquet-Resolved_2026} recently compared
such a static treatment with a secular Floquet--Born--Markov master equation for
two qubits with a harmonically modulated coupling and a parity-breaking
transverse field, and found that the static treatment underestimates the
steady-state entanglement. Our results agree with this conclusion, and we
recover their results when we adopt their model. For the symmetric model
introduced below, and its parity-broken extension, we ask which features of the
steady state are set by symmetry, which by the basis of the dissipator, and
which by the secular approximation, the last by comparing directly with a
nonsecular Floquet master equation.

For our main results, we consider two identical qubits without transverse
fields [Fig.~\ref{fig:model}(a)].
Parity symmetry
divides the Floquet spectrum into two sectors that respond differently to the
drive. 
However, it is not the only symmetry to dictate the selection rule and dynamics, the exchange symmetry, simultaneously, takes role.
As a result of the joint conserved symmetries, the even sector forms avoided crossings at multiphoton resonances, while
the odd sector is exactly solvable and its branches cross. We take the modulation
to be anharmonic, as it is for a nonlinear coupler, so that high-order sidebands
are reached at first order in the drive amplitude. We compare a time-independent
dressed-state master equation (TI-GME) with secular and nonsecular
quasienergy-resolved Floquet master equations (F-GME), which differ in how they
sample the bath [Fig.~\ref{fig:model}(b)].

Our findings fall into two groups. The first concerns the physics of the
driven steady state.
The anharmonic modulation increases the cycle-averaged coupling and thus the ground-state entanglement of the cycle-averaged Hamiltonian. The steady-state concurrence nevertheless stays close to its undriven value and vanishes only at the even-sector anticrossings, as a consequence of a dissipative selection rule.
 The odd crossings, which interleave with them, leave no trace. These
entanglement deaths require the anticrossing gap to exceed the damping rate, and
the resonances are also visible in the emission.

The second concerns the description of the dissipation. Comparing the TI-GME with
secular and nonsecular F-GMEs separates the basis in which the dissipator is built,
static dressed or Floquet, from the secular approximation. The choice of basis
changes the concurrence between the
resonances, by up to $0.2$ for an Ohmic bath and $0.5$ for a flat one, for which
the static description also misses the deaths. The change produced by the secular
approximation in the Floquet basis is orders of magnitude smaller, except near anticrossings whose gap is comparable to the
damping, where the secular F-GME predicts deaths that the damping has washed out.
The choice of basis matters most when the bath has spectral weight at low
frequency, since
the drive moves dissipative channels through zero frequency. Finally, breaking
parity, as in Ref.~\cite{Pacino_Foquet-Resolved_2026}, adds resonances between the
sectors, which explains the richer structure of their maps.

These effects should be accessible in superconducting circuits, for which the
schematic of Fig.~\ref{fig:model} is a representative example. There flux-tunable
couplers are routinely modulated at GHz frequencies to drive parametric
gates~\cite{Niskanen2007,Yan_Tunable_2018,Mitchell_Hardware-Efficient_2021}. The nonlinear response of such a coupler
naturally produces an anharmonic modulation of the coupling, and each qubit
typically has its own loss channel~\cite{Krantz2019}. The relevant scales are
also favorable. For qubit frequencies of a few GHz, the anticrossing gaps found
here are in the MHz range, far above the damping rates of current qubits, so the
resonances can be resolved with ease. The steady state is reached on the scale of
the qubit lifetime, and its $X$-state form allows the concurrence to be extracted
from a few joint populations and coherences.

The rest of the paper is organized as follows. Section~\ref{sec:model} introduces
the model and the cycle-averaged Hamiltonian, and
shows how parity divides the Floquet spectrum into two sectors, and how exchange symmetry produces exact crossings in the odd sector and anticrossings in the even one.
It also derives the selection rule that every
dissipative channel connects the two sectors.
Section~\ref{sec:theory} presents the TI-GME and the secular and nonsecular
F-GMEs, explains how comparing them separates the dissipator basis from the secular
approximation, and defines the baths and observables. Section~\ref{sec:results}
contains the results, namely the spectrum and dissipative channels, the
entanglement deaths and their mechanism, the role of the damping and of the
secular approximation, the maps over drive frequency and the bath dependence, and
finally the harmonic modulation with broken parity.
Section~\ref{sec:discussion} discusses these results and their experimental
realization, and Sec.~\ref{sec:conclusions} concludes. The appendices give the
flat-bath maps (Appendix~\ref{app:flat}), the positivity diagnostics
(Appendix~\ref{app:positivity}), the comparison with
Ref.~\cite{Pacino_Foquet-Resolved_2026} (Appendix~\ref{app:pacino}), and the
Fourier coefficients and the exact odd-sector solution
(Appendix~\ref{app:fourier}).

\section{Model and symmetries}
\label{sec:model}

Figure~\ref{fig:model} shows the model and the two theoretical descriptions of
its dissipation that we compare. 
Two identical qubits interact through a
transverse coupling $g(t)\sigma_x^{(1)}\sigma_x^{(2)}$ that is modulated at
frequency $\omega_d$, and each qubit is damped by its own reservoir
[Fig.~\ref{fig:model}(a)]. This setting is representative of several
experimental platforms.
The most direct one is a pair of
superconducting qubits coupled through a flux-tunable coupler, whose flux is
modulated to activate parametric
interactions~\cite{Niskanen2007,Bialczak2011,Yan_Tunable_2018,Han_ParametricISWAP_2020,Mitchell_Hardware-Efficient_2021,Krantz2019}. Another is a
pair of near-field-coupled emitters whose separation, and hence dipole--dipole
coupling, is modulated~\cite{Ficek2002,Angelatos2015,Chang_Direct_2025}. We consider two forms of
modulation. The anharmonic form, used for the main results, describes a nonlinear
coupler and drives all sidebands at first order. The harmonic form, used in
Sec.~\ref{sec:harmonic}, is the one of Ref.~\cite{Pacino_Foquet-Resolved_2026}.
Each qubit is damped by its own reservoir, with either a flat or an Ohmic spectral
density.

Figure~\ref{fig:model}(b) contrasts the two ways of treating the dissipation. 
The TI-GME builds its jump operators and rates from the static dressed
spectrum, so it samples the bath at a few fixed frequencies. The F-GME obtains
them from the Floquet spectrum, so it samples the bath at a comb of channel
frequencies that move with the drive and can approach zero.
 The rest of this section sets up the
Hamiltonian and its symmetries; Sec.~\ref{sec:theory} gives the master equations.

\subsection{Coupled qubits with a modulated coupling}

The Hamiltonian is ($\hbar=1$)
\begin{equation}
H(t)=\frac{\omega_1}{2}\sigma_z^{(1)}+\frac{\omega_2}{2}\sigma_z^{(2)}
+g(t)\,\sigma_x^{(1)}\sigma_x^{(2)},
\label{eq:H}
\end{equation}
which describes, for example, superconducting qubits coupled inductively or
capacitively~\cite{Krantz2019,Yan_Tunable_2018} or near-field-coupled
emitters~\cite{Ficek2002,Angelatos2015,Chang_Direct_2025}. We take identical
qubits, $\omega_1=\omega_2\equiv\Sigma$, which sets the unit of frequency, and we keep
the counter-rotating terms. For a static coupling $g$ the dimensionless coupling
is $\eta=g/\Sigma$.

The Hamiltonian commutes with the parity operator $\Pi=\sigma_z^{(1)}\sigma_z^{(2)}$ and
splits into an even block, spanned by $\{\ket{gg},\ket{ee}\}$, and an odd block,
spanned by $\{\ket{ge},\ket{eg}\}$. For a static coupling the even block has
energies $\pm\sqrt{\Sigma^2+g^2}$ and the odd block $\pm g$. We label the four
dressed states by energy. The ground state $\ket{G}$ and the doubly excited state
$\ket{T}$ are even, and $\ket{L}$ and $\ket{U}$ are odd, with
$\ket{U},\ket{L}=(\ket{ge}\pm\ket{eg})/\sqrt2$.

The counter-rotating terms mix $\ket{gg}$ and $\ket{ee}$, so the ground state is
entangled, with concurrence~\cite{gunlycke2001,wang2001,arnesen2001}
\begin{equation}
C_G=\frac{\eta}{\sqrt{1+\eta^2}}.
\label{eq:CG}
\end{equation}
For the coupling $\eta=0.5$ used below, $C_G=0.447$.

This ground-state entanglement is carried by virtual excitations, the admixture of
$\ket{ee}$ produced by the counter-rotating terms, in the same way as the virtual
photons in the ground state of ultrastrongly coupled light and
matter~\cite{Ciuti_Quantum_2005,ashhab2010,DeLiberato_Virtual_2017,Stassi_Spontaneous_2013}.
These excitations are not emitted, and they become real only if the coupling is
changed nonadiabatically or a parametric process converts them into radiated
photons~\cite{Stassi_Spontaneous_2013,Kim_Squeezed_2026}.

Equation~\eqref{eq:H} is an effective model. In the platforms of
Fig.~\ref{fig:model}, the coupling is mediated, by a single coupler mode in
circuits or by the near field of the emitters, and each qubit is the two-level
truncation of a multilevel system. At ultrastrong coupling, truncating the
matter breaks gauge invariance unless the coupling is introduced consistently.
Gauge invariance is restored by implementing the minimal-coupling replacement on
the truncated space~\cite{DiStefano_Resolution_2019,Garziano_Gauge_2020,
Savasta_Gauge_2021,Gustin_Gauge_2023}, including for two emitters in a common
cavity~\cite{Akbari_Generalized_2023}. In the dipole gauge the resulting model
couples the truncated coordinates directly, which is the accurate truncation for
strongly anharmonic emitters~\cite{DeBernardis_Breakdown_2018}.

Equation~\eqref{eq:H} has exactly this dipole-gauge form, with each qubit coupled
through $\sigma_x^{(\Lambda)}$, and so do the system--bath couplings introduced in
Sec.~\ref{sec:theory}. The quantities we compute from the full system, namely the
dressed and Floquet spectra, the dissipators built from them, and the emission
rates defined through positive-frequency operators, are therefore gauge
independent~\cite{Settineri_Dissipation_2018,Salmon_Gauge-independent_2022}. We
take the modulation introduced below to enter through the same coupling, as appropriate for a
coupler whose parameter is modulated. What remains gauge dependent is the
definition of the subsystems, and with it the subsystem entanglement, the
virtual-excitation content, and the outcome of time-dependent
interactions~\cite{Settineri_Gauge_2021}. We therefore define the two qubits
operationally, as the systems addressed by local control and readout, and
discuss the meaning of the concurrence in Sec.~\ref{sec:discussion}.

The coupling is modulated periodically, $g(t)=g(t+T)$ with $T=2\pi/\omega_d$. A
tunable coupler is a nonlinear element, so a sinusoidal control signal does not
produce a sinusoidal coupling~\cite{Niskanen2007,Yan_Tunable_2018}. We thus take the
anharmonic form
\begin{equation}
g(t)=\frac{g}{\left[1+r\sin(\omega_d t)\right]^{3}},\qquad 0\le r<1,
\label{eq:gt}
\end{equation}
which is also the coupling of two dipoles whose separation $R$ oscillates with
relative amplitude $r=\delta R/R$~\cite{Chang_Direct_2025,Akbari_Floquet_2025}, i.e., $R\to R(t)=R+\delta R\,\sin(\omega_d t)$. A drive phase would only shift the
time origin and does not affect cycle-averaged quantities, so we set it to zero
throughout.

The Fourier coefficients of Eq.~\eqref{eq:gt}, $g(t)=\sum_m g\,I_m\,e^{-\ii m\omega_d t}$,
are given in closed form in Appendix~\ref{app:fourier}. They decay only
geometrically with $|m|$, so every sideband order is driven at first order in $r$.
A harmonic modulation $g+\delta g\sin\omega_d t$, by contrast, reaches order $m$
only at order $(\delta g)^m$; we return to this in Sec.~\ref{sec:harmonic}.

The cycle average of the coupling,
\begin{equation}
g_0=g\,I_0(r),\qquad I_0(r)=\frac{2+r^2}{2(1-r^2)^{5/2}},
\label{eq:g0}
\end{equation}
exceeds $g$ and grows with $r$. The modulation therefore renormalizes the static
Hamiltonian to $H_0=H|_{g\to g_0}$, whose ground-state concurrence
$C_G^{\rm ren}=C_G|_{g\to g_0}$ increases with $r$. In this regard, the drive makes
the cycle-averaged Hamiltonian more entangling. Starting from the ground state of
$H_0$, the closed-system Floquet dynamics oscillates about this value, and its
long-time average equals $C_G^{\rm ren}$.

We highlight that this renormalization is unique to the anharmonic modulation. For a harmonic
modulation $g+\delta g\sin\omega_d t$ the cycle average is $g$ itself, so the
drive leaves $H_0$ and $C_G$ unchanged. Only the 
\emph{anharmonic} coupling makes the
cycle-averaged Hamiltonian more entangling as the drive grows.

The coupling swings between $\eta/(1+r)^3$ and $\eta_{\max}=\eta/(1-r)^3$. We show
$\eta_{\max}$ on the top axis of the figures and mark $\eta_{\max}=1.3$, close to
the largest normalized coupling reported in superconducting circuits, for a
qubit--oscillator system~\cite{Yoshihara2017}. We use $r\le0.55$. The lowest
resonance discussed below, $l=4$, lies at $\eta_{\max}\approx1.2$. The higher
orders require larger peak couplings, but none of the mechanisms depends on the
size of $\eta$.

\subsection{Parity and exchange symmetry}
\label{sec:symmetry}

The drive multiplies a parity-even operator by a scalar, so $\Pi$ is conserved at
every instant. Each Floquet state therefore carries a parity $\pi_\alpha=\pm1$, the
same for all its sidebands. For identical qubits $H(t)$ also commutes with the
exchange operator $\mathcal{E}$, defined by $\mathcal{E}\ket{ab}=\ket{ba}$. Exchange
acts trivially on the even block but splits the odd block into the triplet
$\ket{U}$ and the singlet $\ket{L}$. As a result the two sectors behave very
differently under the drive.

\emph{Odd sector.} On the odd block $H(t)=g(t)\,\mathcal{E}$, which commutes with
itself at different times. The Floquet states are therefore the static states
$\ket{U}$ and $\ket{L}$ multiplied by a periodic phase, with quasienergies
$\pm g_0$ (mod $\omega_d$) for all $r$ (Appendix~\ref{app:fourier}). The two odd
branches belong to different exchange classes and cross exactly, at
\begin{equation}
2g\,I_0(r)=n\,\omega_d .
\label{eq:oddloci}
\end{equation}

\emph{Even sector.} On the even block the Hamiltonian does not commute with itself
at different times. The Floquet states are genuinely dressed, and the two even
branches, which share one symmetry class, anticross. Each block of $H(t)$ is
traceless, so the quasienergies are $\pm\varepsilon_e$ and $\pm\varepsilon_o$, and
the even anticrossings occur at
\begin{equation}
2\varepsilon_e=l\,\omega_d .
\label{eq:evenloci}
\end{equation}
For $\omega_d=0.6\Sigma$ and $\eta=0.5$, the bound $2\varepsilon_e\ge2\Sigma\sqrt{1+\eta^2}$
implies $l\ge4$.

Table~\ref{tab:anticrossings} lists the positions and gaps of the first even
anticrossings, together with the nearest odd crossings. The two families
interleave closely, only $0.004$ to $0.014$ apart in $r$. This makes it possible
to tell them apart in the steady state.

\emph{Dissipative selection rule.} Each qubit couples to its bath through
$\sigma_x^{(\Lambda)}$, which anticommutes with $\Pi$. Its matrix elements between
Floquet states of equal parity therefore vanish, and every dissipative channel
connects the even and odd sectors. The coherent drive mixes states only within a
sector. The bath therefore sees an even anticrossing through the changed
composition of the even Floquet states it connects to the odd ones. At an odd
crossing the Floquet states do not change, and nothing happens.

Exchange further fixes which odd state is reached. The symmetric combination
$\sigma_x^{(1)}+\sigma_x^{(2)}$ connects only to $\ket{U}$, and the antisymmetric
one only to $\ket{L}$. With independent baths both combinations act with equal
weight, so there is no dark state and the singlet is not protected.

\begin{table}[t]
\caption{Closed-system anticrossings at $\omega_d=0.6\Sigma$, $\eta=0.5$. For the even
sector we give the positions $r^*_l$, the gaps $\delta_l=\min\{2\varepsilon_e,\omega_d-2\varepsilon_e\}$
at $r^*_l$, slopes $s_l=|d(2\varepsilon_e)/dr|$ on either side, and $\delta_l/\gamma$
for $\gamma=10^{-3}\Sigma$. For the odd sector we give the nearest exact crossing
$n$ and its position $r_n$ [Eq.~\eqref{eq:oddloci}].}
\label{tab:anticrossings}
\begin{ruledtabular}
\begin{tabular}{cccccc}
$l$ & $r^*_l$ & $\delta_l/\Sigma$ & $s_l$ & $\delta_l/\gamma$ & $n$, $r_n$ \\
\hline
4 & 0.2479 & $1.30\times10^{-3}$ & 1.53/1.61 & 1.3 & 2, 0.2437 \\
5 & 0.4393 & $4.51\times10^{-3}$ & 5.33/5.69 & 4.5 & 3, 0.4261 \\
6 & 0.5203 & $4.91\times10^{-3}$ & 9.55/9.98 & 4.9 & 4, 0.5100 \\
7 & 0.5706 & $4.56\times10^{-3}$ & 14.2/14.6 & 4.6 & 5, 0.5627 \\
\end{tabular}
\end{ruledtabular}
\end{table}

\section{Open-system theory}
\label{sec:theory}

Each qubit couples to its own reservoir through $X_\Lambda=\sigma_x^{(\Lambda)}$,
$\Lambda=1,2$, as appropriate for circuit qubits with separate loss
channels~\cite{Krantz2019}. Both master equations below follow from the
Born--Markov approximation with the full $H(t)$ in the coherent part. They
differ only in how $X_\Lambda$ is split into components of definite frequency,
which fixes both the jump operators and the frequencies at which the bath is
sampled [Fig.~\ref{fig:model}(b)].

\subsection{Time-independent dressed master equation}

The TI-GME treats the drive as a perturbation of the coherent dynamics only. The
dissipator is derived for the static Hamiltonian $H_0$, with the coupling replaced
by its cycle average $g_0$, and is then kept fixed in time. Building it from the
eigenstates of $H_0$, rather than from the bare qubit states, is essential at
these couplings. A bare-state dissipator would, for example, drive the system
toward $\ket{gg}$ instead of the entangled ground
state~\cite{Settineri_Dissipation_2018,Salmon_Gauge-independent_2022,Akbari_Generalized_2023}.

Writing $H_0\ket{j}=E_j\ket{j}$, the coupling operator splits into components
\begin{equation}
A_\Lambda(\omega)=\sum_{E_k-E_j=\omega}\braket{j|X_\Lambda|k}\,\ket{j}\bra{k},
\end{equation}
with $X_\Lambda=\sum_\omega A_\Lambda(\omega)$. In secular form the master equation
is
\begin{equation}
\dot\rho=-\ii[H(t),\rho]+\sum_{\Lambda}\sum_{\omega}
\gamma_\Lambda(\omega)\,\mathcal{D}[A_\Lambda(\omega)]\rho ,
\label{eq:TIGME}
\end{equation}
with $\mathcal{D}[A]\rho=A\rho A^\dagger-\tfrac12\{A^\dagger A,\rho\}$ and
$\gamma_\Lambda(\omega)=2\,\mathrm{Re}\,\Gamma_\Lambda(\omega)$. The sum runs over all
transition frequencies; at zero temperature the rates vanish for $\omega\le0$.

Equation~\eqref{eq:TIGME} is of Gorini--Kossakowski--Sudarshan--Lindblad (GKLS)
form~\cite{GKS_1976,Lindblad_1976,Davies_Markovian_1974} and preserves positivity. Its rates are evaluated at the static transition
frequencies $\omega_{jk}=E_k-E_j$, which do not depend on $\omega_d$. The drive
enters only through the coherent term, so the TI-GME cannot know that the system
relaxes toward Floquet states rather than toward eigenstates of $H_0$.

\subsection{Floquet master equation}

The F-GME instead builds the dissipator from the Floquet states
(drive dependent). By Floquet's
theorem the solutions of the Schr\"odinger equation with $H(t)$ take the form
$e^{-\ii\varepsilon_\alpha t}\ket{\alpha(t)}$, with periodic Floquet modes
$\ket{\alpha(t)}=\ket{\alpha(t+T)}$ and quasienergies $\varepsilon_\alpha$ defined
modulo $\omega_d$~\cite{Shirley_Solution_1965,Sambe_Steady_1973}. We obtain them
numerically and expand the modes in Fourier harmonics.

In this basis, the coupling operator splits as~\cite{Grifoni_Driven_1998,
Kohler_FloquetMarkov_1997,blumel1991,breuer2000,szczygielski2013,Akbari_Floquet-Liouville_2026,Akbari_Floquet_2026}
\begin{align}
X_\Lambda&=\sum_{\alpha\beta l}A_\Lambda(\Delta_{\alpha\beta l};t),\\
A_\Lambda(\Delta_{\alpha\beta l};t)&=X^\Lambda_{\alpha\beta l}\,e^{-\ii l\omega_d t}
\ket{\alpha(t)}\bra{\beta(t)},
\end{align}
where $X^\Lambda_{\alpha\beta l}$ is the $l$th Fourier component of
$\braket{\alpha(t)|X_\Lambda|\beta(t)}$. Each term is a dissipative channel at
frequency
\begin{equation}
\Delta_{\alpha\beta l}=\varepsilon_\beta-\varepsilon_\alpha+l\,\omega_d ,
\label{eq:Delta}
\end{equation}
where the index $l$ counts the drive quanta exchanged in the transition. A transition
between the same pair of states can therefore emit into the bath at many
frequencies, with weights $|X^\Lambda_{\alpha\beta l}|^2$ set by the dressing.

In nonsecular (Redfield) form, the master equation reads
\begin{multline}
\dot\rho=-\ii[H(t),\rho]+\sum_{\Lambda}\sum_{\Delta,\Delta'}
\Big\{\Gamma_\Lambda(\Delta)\big[A\rho A'^\dagger-A'^\dagger A\rho\big]\\
+\Gamma^*_\Lambda(\Delta')\big[A\rho A'^\dagger-\rho A'^\dagger A\big]\Big\},
\label{eq:FGME}
\end{multline}
with $A\equiv A_\Lambda(\Delta;t)$ and $A'\equiv A_\Lambda(\Delta';t)$. The
cross terms with $\Delta\neq\Delta'$ couple populations and coherences of the
Floquet states. This generator is not of GKLS form, and its positivity has to be
checked (Appendix~\ref{app:positivity}). Partial-secular and related
constructions restore complete positivity at the cost of some
accuracy~\cite{Farina_Open_2019,Nathan_Universal_2020,Mozgunov_Completely_2020,
Trushechkin_Unified_2021}.

Keeping only the terms with $\Delta=\Delta'$ gives the secular F-GME. It is of GKLS
form in the basis of the Floquet modes and reduces to rate equations for the
Floquet populations~\cite{Breuer_Theory_2002,Grifoni_Driven_1998,blumel1991,breuer2000,szczygielski2013}. This
approximation requires $\gamma\ll\delta^*$, where $\delta^*$ is the smallest
spacing between distinct channel frequencies. Near an anticrossing $\delta^*$
becomes comparable to the gap, so this condition can fail exactly where the
interesting physics happens, a general difficulty of near-degenerate Floquet
spectra~\cite{Hone_Statistical_2009}.

The TI-GME and F-GME therefore differ in two ways. The F-GME samples the bath at
the channels $\Delta_{\alpha\beta l}$, which move with $r$ and $\omega_d$ and can
approach zero. Its jump operators also connect Floquet states, which differ from
the static dressed states near an anticrossing. We refer to these two differences
together as the \emph{basis} of the dissipator, static dressed for the TI-GME and Floquet
for the F-GME.

We separate the two choices by comparing pairs of equations that differ in only
one of them~\cite{cattaneo2019,hofer2017}. The \emph{secular shift} of an equation is the change produced by the
secular approximation, $\Delta\Cbar_{\rm sec}=|\Cbar_{\rm sec}-\Cbar_{\rm nonsec}|$,
for the F-GME or for the TI-GME. It measures
the effect of the approximation relative to the nonsecular Born--Markov result,
which contains the secular one as a limit; it is not an error relative to an
exact solution, which is not available. The \emph{basis shift} is the difference
between the F-GME and the TI-GME when both are secular,
$\Delta\Cbar_{\rm basis}=|\Cbar_{\rm F}-\Cbar_{\rm TI}|$, and isolates the basis of
the dissipator.

\begin{figure*}[t]
\centering
\includegraphics[width=\textwidth]{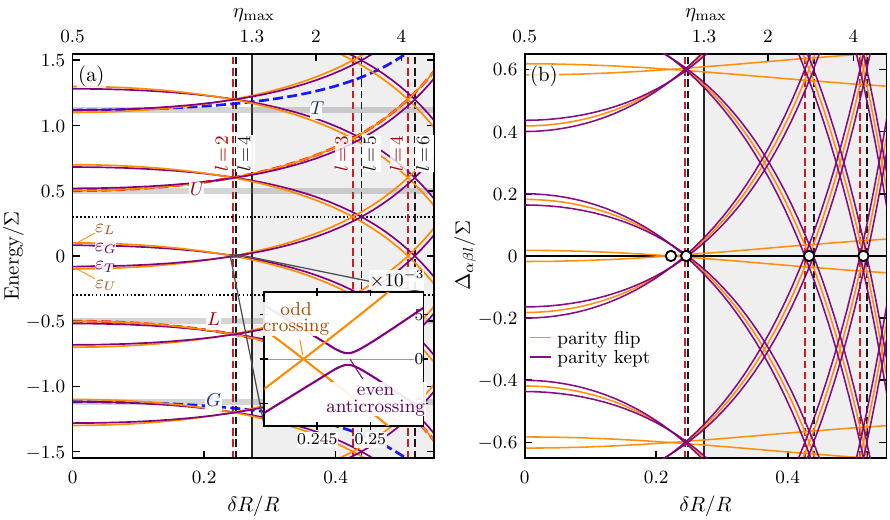}
\caption{Closed-system spectrum and Floquet channels versus modulation depth
$r=\delta R/R$, for $\eta=0.5$ and $\omega_d=0.6\Sigma$.
The top axes show the peak coupling $\eta_{\max}=\eta/(1-r)^3$, and the vertical
line and shading mark $\eta_{\max}>1.3$.
(a)~Floquet quasienergies of even (purple) and odd (orange) parity, with the
static dressed levels (gray bands) and the levels of the cycle-averaged
Hamiltonian $H_0$ (dashed curves, blue for the even levels $G$ and $T$ and red
for the odd levels $L$ and $U$). Dashed vertical guides mark the even anticrossings
($l=4,5,6$, black) and the odd crossings ($l=2,3,4$, red) of
Table~\ref{tab:anticrossings}.
The inset shows that near $\varepsilon=0$ the odd branches cross exactly, while
the even branches anticross with a gap of $1.3\times10^{-3}\Sigma$.
(b)~Channel frequencies $\Delta_{\alpha\beta l}$ for all pairs of states and numbers of drive quanta (harmonics) $l$, colored by whether the transition flips parity (orange, the
dissipative channels) or keeps it (purple). Open circles mark where dissipative
channels pass through zero frequency. The first, at $r\simeq0.2225$, produces the
step in the flat-bath results of Fig.~\ref{fig:main}.}
\label{fig:spectrum}
\end{figure*}

\subsection{Baths and observables}

At zero temperature and neglecting Lamb shifts~\cite{gramich2014}, the bath correlation spectrum is
$\Gamma_\Lambda(\omega)=(\gamma/2)\,f(\omega)\,\Theta(\omega)$, with $\Theta(0)=0$, so
channels at zero frequency do not dissipate. We use $f=1$ for a flat
bath and $f=\omega/\Sigma$ for an Ohmic bath. The two baths agree at the qubit
frequency, and only channels with $\Delta>0$ dissipate. The flat bath has a step
at $\omega=0$, so its rate switches on or off when a channel crosses zero
frequency. The Ohmic rate instead vanishes continuously.

\begin{figure*}[t]
\centering
\includegraphics{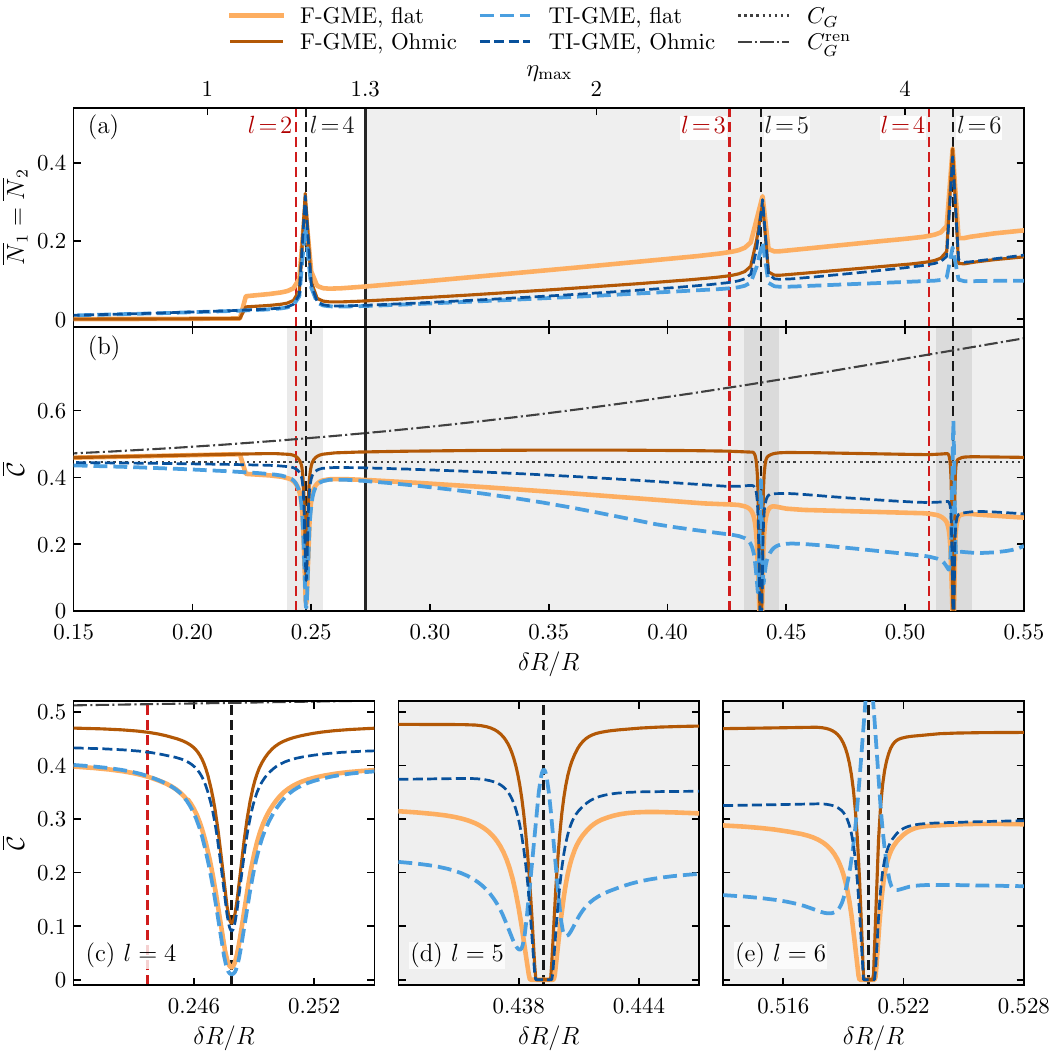}
\caption{Cycle-averaged steady-state population and concurrence versus $r$, for
the parameters of Fig.~\ref{fig:spectrum}, $\gamma=10^{-3}\Sigma$ and $T=0$, for a
flat bath (light curves) and an Ohmic bath (dark curves). Solid lines show the
nonsecular F-GME and dashed lines the secular TI-GME.
(a)~Population $\Nbar_1=\Nbar_2$.
(b)~Concurrence $\Cbar$, with the ground-state concurrences of the static
($C_G$, dotted) and cycle-averaged ($C_G^{\rm ren}$, dash-dotted) Hamiltonians;
shaded bands mark the windows magnified in (c)--(e). Dashed vertical guides mark
the even anticrossings (black) and odd crossings (red).
(c)--(e)~$\Cbar$ near the even resonances $l=4$, $5$, $6$.}
\label{fig:main}
\end{figure*}

We take $\gamma_1=\gamma_2=\gamma=10^{-3}\Sigma$ unless stated otherwise, so the
damping is weak compared with all static scales but comparable to the
anticrossing gaps of Table~\ref{tab:anticrossings}. The Ohmic bath, which is
closer to the environments of circuit qubits, is used for the main results and
maps (Figs.~\ref{fig:main} and~\ref{fig:maps}), with the flat bath shown
alongside in Fig.~\ref{fig:main}. The flat bath shows the largest effect of the dissipator basis and
is used for the damping and secular tests of Figs.~\ref{fig:gamma}
and~\ref{fig:secular}; for the Ohmic bath we summarize these tests in the text.

We compute the time-periodic steady state $\rho_{\rm ss}(t)$ directly in
Floquet--Liouville space~\cite{Akbari_Floquet-Liouville_2026} and report cycle
averages, denoted by an overbar. The population of qubit~1 is
$\Nbar_1=\overline{\mathrm{Tr}[X_1^-X_1^+\rho_{\rm ss}(t)]}$, where $X_1^+$ is the
positive-frequency part of $X_1$. We decompose it in the Floquet states for the
F-GME and in the static dressed states for the TI-GME, consistent with the
respective dissipators.

We quantify the entanglement by the cycle-averaged
concurrence~\cite{Hill_Entanglement_1997,Wootters_Entanglement_1998,Horodecki_Quantum_2009}
\begin{equation}
\Cbar=\frac1T\int_0^T C[\rho_{\rm ss}(t)]\,\dd t ,
\end{equation}
evaluated in the bare basis, that is, relative to the partition into the two
operationally defined qubits (note $\Cbar\neq C[\overline{\rho_{\rm ss}(t)}]$). At each
time,
\begin{equation}
C[\rho]=\max\{0,\lambda_1-\lambda_2-\lambda_3-\lambda_4\},
\label{eq:Wootters}
\end{equation}
where $\lambda_i$ are the square roots of the eigenvalues of
$\rho\,(\sigma_y\otimes\sigma_y)\,\rho^*\,(\sigma_y\otimes\sigma_y)$ in decreasing
order~\cite{Wootters_Entanglement_1998}.

For the symmetric model, parity conservation makes $\rho_{\rm ss}(t)$ an $X$
state, and exchange symmetry gives $\rho_{ge,ge}=\rho_{eg,eg}=P_{o}/2$, with $P_o$ the
odd-sector population. The concurrence then reduces to
\begin{equation}
C=2\max\Big\{0,\ |\rho_{gg,ee}|-\tfrac{P_o}{2},\
|\rho_{ge,eg}|-\sqrt{\rho_{gg,gg}\,\rho_{ee,ee}}\Big\}.
\label{eq:Cx}
\end{equation}

At zero temperature the odd coherence $\rho_{ge,eg}$ is negligible. Entanglement
then requires the pair coherence $|\rho_{gg,ee}|$ to exceed half the odd
population. For the dressed ground state $|\rho_{gg,ee}|=C_G/2$ and $P_o=0$.

All concurrences shown are computed from the general expression,
Eq.~\eqref{eq:Wootters}. For the symmetric model it reduces exactly to Eq.~\eqref{eq:Cx}, which we use to interpret the results. When parity is broken (Sec.~\ref{sec:harmonic} and
Appendix~\ref{app:pacino}), the steady state is no longer of $X$ form and only the general expression applies.

\section{Results}
\label{sec:results}

\subsection{Floquet spectrum and dissipative channels}

Figure~\ref{fig:spectrum}(a) shows the quasienergies at $\omega_d=0.6\Sigma$. The
odd branches are replicas of the static odd levels at the renormalized coupling
and cross one another freely. The even branches form anticrossings at the
positions of Table~\ref{tab:anticrossings}, with gaps of a few $10^{-3}\Sigma$
(inset).

Figure~\ref{fig:spectrum}(b) shows the channel frequencies. By the selection rule,
only the parity-flipping (orange) channels carry dissipation. Several of them pass
through zero frequency as $r$ grows. The first does so at $r\simeq0.2225$, the
next at $r\simeq0.246$, close to the $l=4$ resonance, and others near $l=5$
and $6$ (open circles). These zero crossings
matter for a bath whose spectral density is finite at $\omega\to0$, as discussed
in Sec.~\ref{sec:bath}.

\begin{figure*}[t]
\centering
\includegraphics{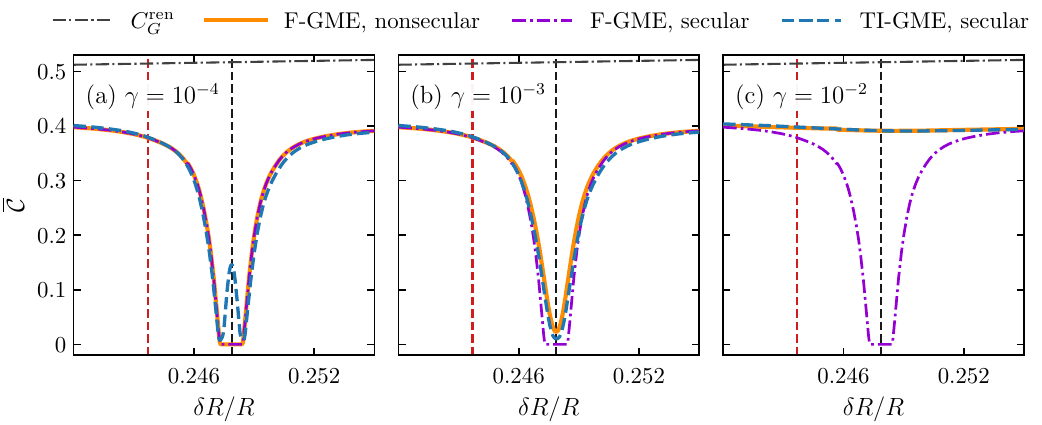}
\caption{Concurrence in the $l=4$ window (flat bath) for
(a)~$\gamma=10^{-4}\Sigma$, (b)~$10^{-3}\Sigma$ and (c)~$10^{-2}\Sigma$, from the
nonsecular F-GME (solid), the secular F-GME (dash-dotted) and the secular TI-GME
(dashed); the dash-dotted gray line is $C_G^{\rm ren}$. Vertical guides mark the
even anticrossing (black) and the nearest odd crossing (red). The gap is
$\delta_4=1.3\times10^{-3}\Sigma$.}
\label{fig:gamma}

\vspace{12pt}
\centering
\includegraphics[width=0.97\textwidth]{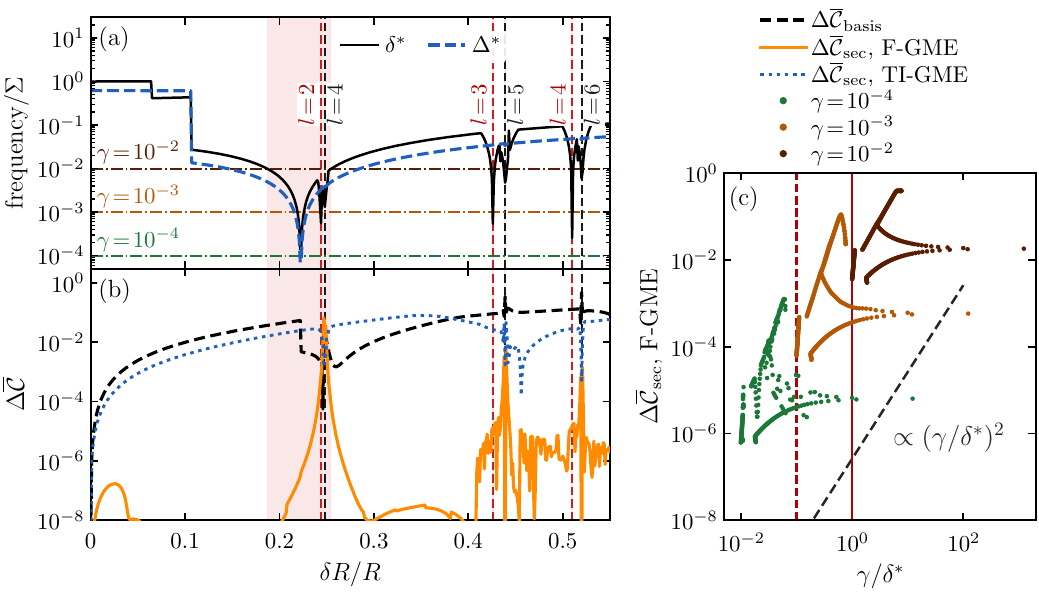}
\caption{Validity of the secular approximation as a function of $r$, at
$\omega_d=0.6\Sigma$ and $\eta=0.5$ as in
Fig.~\ref{fig:main}, for the flat bath.
(a)~Smallest channel frequency $\Delta^*=\min|\Delta_{\alpha\beta l}|$ (blue
dashed) and smallest spacing $\delta^*$ between distinct channels connected by the
bath (black); horizontal lines mark $\gamma/\Sigma=10^{-4}$, $10^{-3}$, $10^{-2}$.
The shading marks where $\gamma/\delta^*>0.1$ for $\gamma=10^{-3}\Sigma$.
(b)~Secular shifts $\Delta\Cbar_{\rm sec}$ of the F-GME (orange) and of the TI-GME
(blue dotted), and basis shift $\Delta\Cbar_{\rm basis}$ between the secular F-GME and
TI-GME (black dashed), shown down to $10^{-8}$ (definitions in
Sec.~\ref{sec:theory}).
(c)~Secular shift $\Delta\Cbar_{\rm sec}$ of the F-GME versus $\gamma/\delta^*$ in the $l=4$ window for
$\gamma/\Sigma=10^{-4}$, $10^{-3}$, $10^{-2}$. Red lines mark
$\gamma/\delta^*=0.1$ (dashed) and $1$ (solid); the dashed black line is a
$(\gamma/\delta^*)^2$ guide.}
\label{fig:secular}
\end{figure*}

\subsection{Entanglement death at the even resonances}

Figure~\ref{fig:main} shows the steady state as a function of the modulation
depth $r$, at the same drive frequency $\omega_d=0.6\Sigma$ and coupling
$\eta=0.5$ as in Fig.~\ref{fig:spectrum}. The concurrence
does not follow $C_G^{\rm ren}$, which rises from $0.47$ to $0.82$ over the range
shown. With the Ohmic bath the F-GME concurrence stays close to its undriven
value, between $0.46$ and $0.48$. The TI-GME concurrence instead decreases
steadily, to $0.29$ at $r=0.55$. With the flat bath both are lower
(Sec.~\ref{sec:bath}).

On top of this background the F-GME concurrence vanishes at the even resonances
[Figs.~\ref{fig:main}(c)--\ref{fig:main}(e)]. At $l=5$ and $6$ it is exactly zero
over a finite window, for both baths. At $l=4$ it drops to $0.02$ (flat) and
$0.11$ (Ohmic) without reaching zero, since the gap there is only $1.3\gamma$
(Table~\ref{tab:anticrossings} and Sec.~\ref{sec:damping}). The population peaks
at the same positions.

No feature appears at the odd crossings, although each lies only $0.004$--$0.014$
from an even resonance. Every resonant feature in $\Nbar$ and $\Cbar$ lies within
half a grid step of an even anticrossing. This is the selection rule of
Sec.~\ref{sec:symmetry} at work.

\begin{figure*}[t]
\centering
\includegraphics{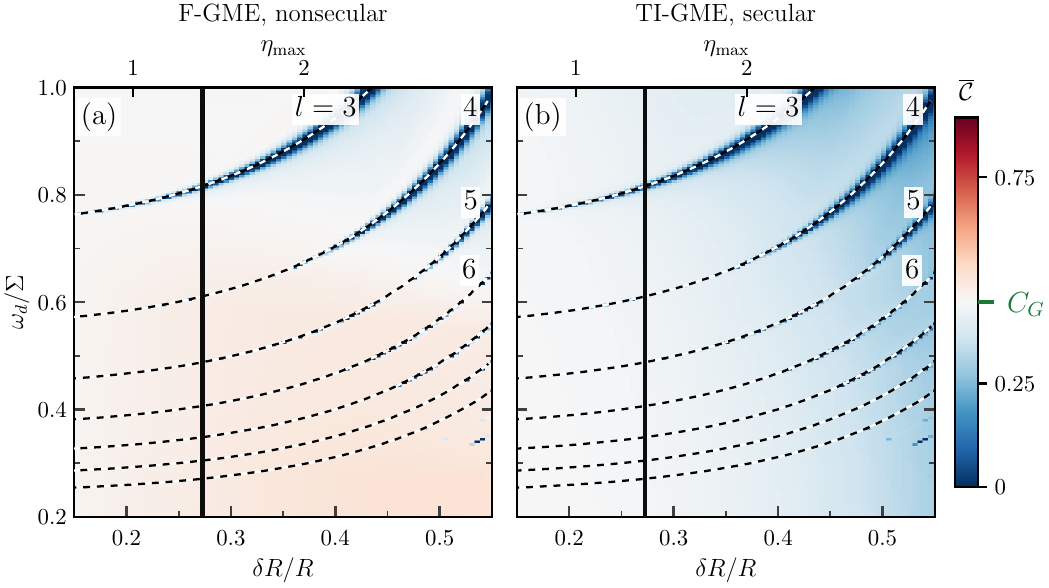}
\caption{Concurrence maps versus $(r,\omega_d)$ for the Ohmic bath,
$\gamma=10^{-3}\Sigma$, for (a)~the nonsecular F-GME and (b)~the secular TI-GME, with the
even-sector loci $2\varepsilon_e=l\omega_d$ (dashed). The color scale is centered
on $C_G$; the vertical line marks $\eta_{\max}=1.3$.}
\label{fig:maps}
\end{figure*}

Equation~\eqref{eq:Cx} explains the deaths. Away from resonance the dissipator
returns the system toward the ground-state Floquet mode, which retains a pair
coherence close to $C_G/2$. The small odd population produced by the drive then
leaves $\Cbar$ near $C_G$. At an even anticrossing the two even Floquet states
become equal superpositions of $\ket{G}$ and $\ket{T}$.

The steady state then carries the pair coherences of $\ket{G}$ and $\ket{T}$, which
have opposite signs, with comparable weights, so $|\rho_{gg,ee}|$ collapses. At
the same time, decay out of the $\ket{T}$ component feeds the odd population.
Once $|\rho_{gg,ee}|\le P_o/2$ throughout the cycle, $\Cbar=0$. This requirement is
stronger than a zero of the concurrence at a single phase of the cycle.

The two master equations predict the resonances at the same positions, but not
always with the same form.
With the flat bath the TI-GME does not produce the deaths at $l=5$ and
$6$; its concurrence reaches only $0.06$ and $0.12$ and shows dispersive features,
overshooting to $0.55$ next to $l=6$.

With the Ohmic bath the TI-GME reproduces the deaths at $l=5$ and $6$, but on a
background lower than the F-GME one by $0.10$ on average and by up to $0.18$. In
both cases the TI-GME concurrence lies below the F-GME one away from the
resonances. This agrees with the finding of Ref.~\cite{Pacino_Foquet-Resolved_2026}
that the static treatment underestimates the entanglement.

\begin{figure*}[t]
\centering
\includegraphics[width=\textwidth]{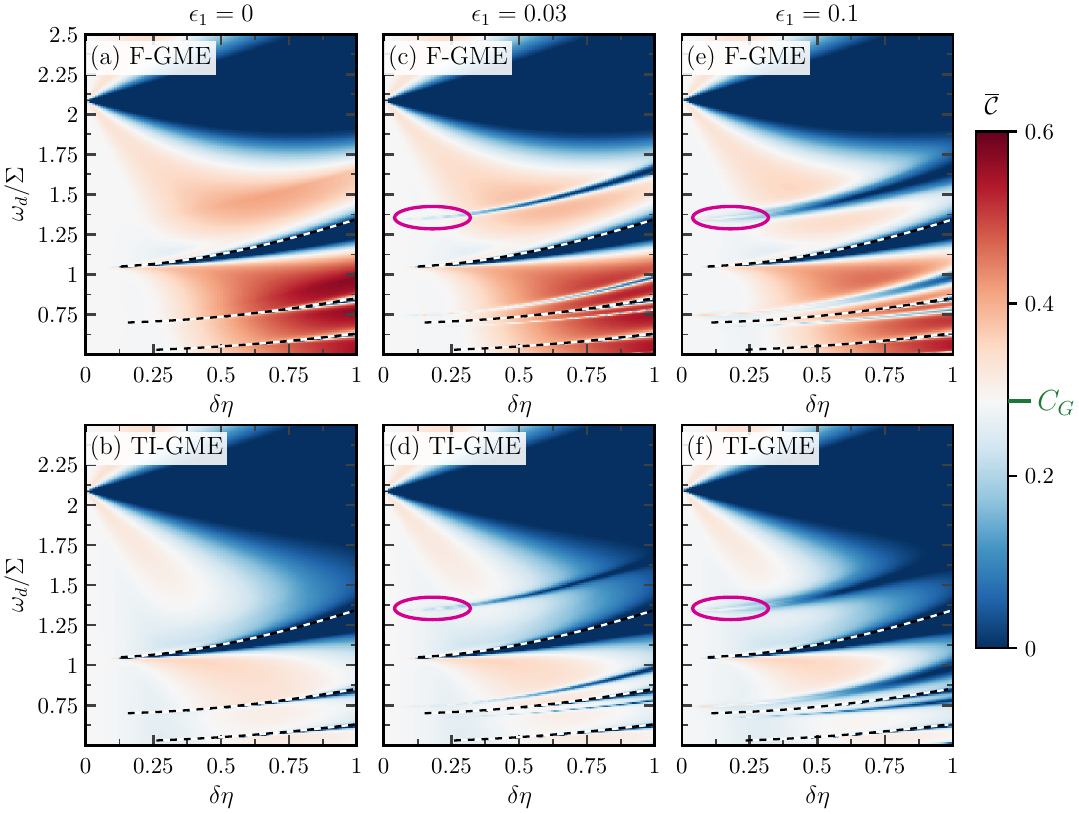}
\caption{Concurrence maps for the harmonic modulation
$g(t)=\Sigma\,[\eta+\delta\eta\sin(\omega_d t)]$ with $\eta=0.3$, an Ohmic bath and
$\gamma=10^{-3}\Sigma$, versus $(\delta\eta,\omega_d)$. Panels (a), (c) and (e) show the
nonsecular F-GME and panels (b), (d) and (f) the secular TI-GME. The columns correspond to a
transverse field $\epsilon_1=0$, $0.03\Sigma$ and $0.1\Sigma$ on qubit~1
($\epsilon_2=0$). The color scale is centered on
$C_G=\eta/\sqrt{1+\eta^2}\simeq0.287$. Dashed lines mark the even-sector
resonances $2\varepsilon_e=l\omega_d$ of the symmetric problem ($\epsilon_1=0$),
repeated in the other columns for reference. 
The features that appear away from them for $\epsilon_1\neq0$ are resonances
between the even and odd sectors activated by the transverse field; magenta
ellipses mark the onset of the clearest one, near $\omega_d\simeq1.34\Sigma$.
}
\label{fig:harmonic}
\end{figure*}

\subsection{Damping and the secular approximation}
\label{sec:damping}

The deaths are resolved only if the anticrossing gap exceeds the damping rate.
Figure~\ref{fig:gamma} shows the $l=4$ window, where $\delta_4=1.3\times10^{-3}\Sigma$,
for three damping rates. At $\gamma=10^{-4}\Sigma$ both F-GMEs give $\Cbar=0$ at the
anticrossing. The TI-GME gives a dip with a central peak, reaching $\Cbar=0.15$ at
$r^*_4$.

At $\gamma=10^{-3}\Sigma\approx\delta_4$ the nonsecular F-GME no longer reaches zero,
with a minimum of $0.02$. At $\gamma=10^{-2}\Sigma$ the dip is washed out into a shallow, broad depression,
and the nonsecular F-GME and the TI-GME agree, both giving $\Cbar\approx0.39$ at the
anticrossing.
 The secular
F-GME, by contrast, still predicts complete entanglement death at
$\gamma=10^{-2}\Sigma$. It assumes that the damping resolves channels which, at
this rate, it does not, and is wrong here.

Figure~\ref{fig:secular} quantifies this over the full range of $r$ of
Fig.~\ref{fig:main}, at $\omega_d=0.6\Sigma$. The secular shift of the F-GME is
below $5\times10^{-7}$ away from the resonances. It grows where $\gamma/\delta^*$
exceeds about $0.1$, reaching $7\times10^{-2}$ at $l=4$, and $7\times10^{-3}$ and
$1.5\times10^{-3}$ at $l=5$ and $6$~\cite{eastham2016}. It scales as $(\gamma/\delta^*)^2$ before
saturating for $\gamma/\delta^*\gtrsim1$ [Fig.~\ref{fig:secular}(c)].

The basis shift is much larger, $0.03$ (median) between the resonances and up to
$0.35$ and $0.48$ at $l=5$ and $6$. The disagreement between the two descriptions
therefore comes from the basis of the dissipator, not from the secular
approximation. In the Floquet basis the secular approximation is reliable
everywhere except within the narrow windows where $\gamma\gtrsim\delta^*$.

The static basis behaves differently. There the secular shift of the TI-GME is
$0.02$ (median) and up to $0.08$, comparable to the basis shift itself
[Fig.~\ref{fig:secular}(b)]. The drive couples the static dressed states, so
their transition frequencies are not the ones the bath resolves, and dropping
the cross terms between them is not justified.
The secular approximation is therefore safe only in the basis adapted to the
drive. In the Floquet basis the dissipative channels are well separated, except
near anticrossings, so the cross terms between them average out over the
damping time and can be dropped. In the static basis the drive keeps mixing the
static dressed states, so the cross terms do not average out, and dropping them
changes the steady state.

The Ohmic bath gives a similar picture, with one exception.
The secular shift of the F-GME is
below $3\times10^{-7}$ between the resonances and $1.6\times10^{-2}$ and
$1.2\times10^{-2}$ at $l=5$ and $6$, while the basis shift reaches $0.18$ and the
secular shift of the TI-GME $0.32$. The exception is the $l=4$ window, where
$\gamma\approx\delta_4$ and the secular shift of the F-GME, $0.18$, is comparable to
the basis shift there.

\subsection{Dependence on the drive frequency and on the bath}
\label{sec:bath}

Figure~\ref{fig:maps} extends these results from the single drive frequency
$\omega_d=0.6\Sigma$ to drive frequencies between $0.2\Sigma$ and
$\Sigma$. Each resonance order $l$ traces an arc $2\varepsilon_e(r,\omega_d)=l\omega_d$.
The arc exists only for $\omega_d\gtrsim2\Sigma\sqrt{1+\eta^2}/l$ and moves to
larger $r$ as $\omega_d$ increases.

In both panels the concurrence vanishes along these arcs and nowhere else. The
deaths appear as the dark-blue bands that follow the dashed lines labeled
$l=3$--$6$, most clearly toward the upper right of each panel, where the
higher-order arcs have the largest gaps.
Away from the arcs, panel (a) is pale or slightly red, at or above $C_G$, whereas
panel (b) turns blue toward large $r$, below $C_G$.
The odd-crossing loci of Eq.~\eqref{eq:oddloci} again leave no mark. Between the
arcs the F-GME concurrence stays at or slightly above $C_G$, up to $0.51$, whereas
the TI-GME concurrence never exceeds $C_G$ and decreases toward large $r$. The
higher-order arcs are narrower and are resolved only where their gaps exceed
$\gamma$.

The bath spectrum enters through the channels that cross zero frequency
[Fig.~\ref{fig:spectrum}(b)]. For a flat bath the rate of such a channel jumps
between $0$ and $\gamma$ when its frequency changes sign. This produces the step in
the F-GME results~\cite{engelhardt2019} at $r\simeq0.2225$ in Fig.~\ref{fig:main} and the sharp edges
across the flat-bath maps of Appendix~\ref{app:flat}. For an Ohmic bath the rate
vanishes linearly at zero frequency and the steady state is continuous.

The effect is not confined to these edges. The two baths weight channels near zero
frequency very differently, and the F-GME concurrence at $r=0.55$ is $0.28$ for the
flat bath against $0.46$ for the Ohmic one. The TI-GME has no near-zero channels,
and its dependence on the bath is weaker, $0.20$ against $0.29$. The Floquet
steady state therefore depends on the low-frequency part of the bath spectrum,
which a static treatment does not resolve.

With Floquet-resolved operators, the population $\Nbar_1$ shows a small step at
$r\simeq0.2225$ even for the Ohmic bath [Fig.~\ref{fig:main}(a)]. This reflects a
component of $X_1$ changing the sign of its frequency, not a change in the state. Indeed, in the concurrence, the discontinuity of the states is the only effective one, and the discontinuity of the operator decomposition with respect to the observable channels is absent, unlike the populations.

\subsection{Harmonic modulation and broken parity}
\label{sec:harmonic}

To connect with Ref.~\cite{Pacino_Foquet-Resolved_2026} we turn to the harmonic
modulation $g(t)=\Sigma[\eta+\delta\eta\sin\omega_d t]$ and add a transverse field,
$H\to H+\tfrac12\epsilon_1\sigma_x^{(1)}$, which breaks both the parity and the exchange symmetry.
The steady state is then no longer an $X$ state, and $\Cbar$ is computed from
the general expression, Eq.~\eqref{eq:Wootters}.
This is the model of Ref.~\cite{Pacino_Foquet-Resolved_2026}, in which the coupling
is $g_0+A\cos(\omega_d t)$ and frequencies are measured in units of the qubit
frequency $\Delta$ (our $\Sigma$). Our $\delta\eta$ therefore corresponds to their
$A/\Delta$, and their cosine modulation differs from ours only by a shift of the
time origin, which does not affect cycle averages.
Figure~\ref{fig:harmonic} shows the concurrence for $\epsilon_1=0$, $0.03\Sigma$ and
$0.1\Sigma$, the last being the value of Ref.~\cite{Pacino_Foquet-Resolved_2026}.

For $\epsilon_1=0$ the structure is that of the anharmonic case. Only the
even-sector resonances appear, as tongues that open from the
diabatic resonance frequencies and bend toward larger $\omega_d$ as $\delta\eta$
grows. With $\epsilon_1\neq0$ every pair of states can mix, and additional
features appear away from these arcs, more strongly for larger $\epsilon_1$. The
clearest, near $\omega_d\simeq1.34\Sigma=(\sqrt{1+\eta^2}+\eta)\Sigma$, is a
one-photon resonance between an even and an odd level, which parity forbids at
$\epsilon_1=0$.

The maps of Ref.~\cite{Pacino_Foquet-Resolved_2026} therefore contain resonances
that are absent in the symmetric problem, and the selection rule of
Sec.~\ref{sec:symmetry} is a property of the symmetric point. With their model and
settings, our implementation of the two methods reproduces their populations and
concurrence (Appendix~\ref{app:pacino}).

\section{Discussion}
\label{sec:discussion}

Comparing static and Floquet treatments involves two choices that are easily
conflated, the basis of the dissipator and the secular approximation. Within the
Floquet basis the secular approximation is accurate except in windows of width
$\sim\delta_l/s_l\sim10^{-3}$ in $r$ around anticrossings with
$\gamma\gtrsim\delta_l$, whereas changing the basis changes $\Cbar$ throughout. In
the static basis the secular approximation is itself unreliable, changing $\Cbar$
by $10^{-2}$ to $10^{-1}$, so a static treatment cannot be rescued by dropping or
keeping the cross terms. The two equations also differ in positivity. The secular
TI-GME is of GKLS form and gives positive states, while the nonsecular F-GME is
not, but its steady states are physical to within $1.3\times10^{-7}$ over the
whole map of Fig.~\ref{fig:maps} (Appendix~\ref{app:positivity}).

The physical picture that emerges depends partly on the form of the drive. The
renormalization of the cycle-averaged coupling, and with it the more entangling
effective Hamiltonian, is unique to the anharmonic modulation, since a harmonic
modulation leaves the cycle average unchanged. The contrast between a rising
$C_G^{\rm ren}$ and a steady-state concurrence that stays near its undriven value
is therefore specific to nonlinear couplers. The selection rule and the
dependence on the dissipator basis do not rely on it, as the harmonic results of
Sec.~\ref{sec:harmonic} show.

The vanishing of $\Cbar$ resembles entanglement sudden death and
revival~\cite{YuEberly2004,YuEberly_Science_2009,FicekTanas2006,Almeida2007},
but differs in three respects. It occurs in the nonequilibrium steady state as a
function of a control parameter rather than in time. It is caused by
drive-induced anticrossings rather than by the decay of an initially entangled
state. Its appearance also depends on how the dissipator resolves the Floquet
spectrum.

At ultrastrong coupling it is important to be clear about what the concurrence
means. It quantifies the entanglement of the two qubits as subsystems defined by
local control and readout, and it has a direct operational meaning through a
sudden switch-off of the coupling~\cite{Settineri_Gauge_2021}. If the coupling is
switched off nonadiabatically at some phase of the cycle, the state is left
unchanged, but the bare states become the energy eigenstates and the two qubits
are uncoupled and uniquely defined. Virtual excitations, including the $\ket{ee}$
admixture of the ground state, become real and are emitted into the output lines
of the two qubits~\cite{Stassi_Spontaneous_2013,Garziano_Switching_2013}. Local
readout, or photodetection and homodyne detection of the two outputs, then
measures the populations and coherences entering Eq.~\eqref{eq:Cx}. In this sense
$C[\rho_{\rm ss}(t)]$ is the entanglement found after switching off the coupling
at time $t$, and a vanishing $\Cbar$ means that no entanglement is found
whichever phase is chosen. For a coupler the switch-off is a physical change of a
circuit parameter, and a gauge-invariant description of the time-dependent
coupling fixes its outcome~\cite{Settineri_Gauge_2021}. 
For the symmetric model the steady state is an $X$ state, so the measurement
requires only four joint populations and two coherence magnitudes.


The deaths also have signatures that do not rely on this partition. By
Eq.~\eqref{eq:Cx}, $\Cbar$ vanishes when the pair coherence falls below half the
odd population. The odd population is carried by emitting states, and its rise
at the even resonances appears in the emission rates $\Nbar_\Lambda$, which we
define through positive-frequency operators as appropriate for the detected
output at ultrastrong coupling~\cite{Gardiner_Input_1985,Ciuti_InputOutput_2006,
Ridolfo_Photon_2012,Salmon_Gauge-independent_2022}. The peaks of $\Nbar_1$ in
Fig.~\ref{fig:main}(a) coincide with the deaths, so the resonance positions and
the selection rule can be tested by spectroscopy of the emitted fields alone. A
more direct signature would be the two-photon correlation
$\langle X_1^+X_2^+\rangle$ of the fields emitted by the two qubits, the output
counterpart of the pair coherence, which is accessible to cross-correlated
homodyne detection of the two output lines~\cite{Pacino_Foquet-Resolved_2026}.
Unlike $\rho_{gg,ee}$, it vanishes in the ground state, so it records only real,
emitted correlations.


Our predictions are within reach of current superconducting circuits. The
model maps directly onto two qubits coupled through a flux-tunable
coupler~\cite{Krantz2019,Yan_Tunable_2018}. Modulating the coupler flux at $\omega_d$ gives
a periodic coupling whose higher harmonics are set by the nonlinearity of the
coupler, which is the situation captured by Eq.~\eqref{eq:gt}. For
$\Sigma/2\pi=5$~GHz, the drive frequency used here is $\omega_d/2\pi=3$~GHz,
within the range of parametric gates, and the gaps of
Table~\ref{tab:anticrossings} correspond to $\delta_l/2\pi\approx6$--$25$~MHz.

Qubit lifetimes of tens of microseconds give $\gamma/\Sigma\sim10^{-6}$, far
below these gaps, so experiments would operate in the regime of
Fig.~\ref{fig:gamma}(a), where the deaths are sharply resolved and the secular
F-GME is accurate. Adding controlled loss, for example through a Purcell-filtered
readout resonator, would allow $\gamma$ to be tuned across $\delta_l$ and the
crossover of Fig.~\ref{fig:gamma} to be tested. Filtering the environment would
similarly change the low-frequency spectral density and probe the bath dependence
of Sec.~\ref{sec:bath}. The coupling $\eta=0.5$ is at the upper end of what is
currently achievable for two qubits. For smaller couplings the even resonances
occur at higher order and with smaller gaps, so the selection rule and the
dependence on the dissipator basis persist, but resolving them requires weaker
damping, which is the natural regime of superconducting qubits.

Several extensions remain open. The selection rule relies on parity, which
survives qubit detuning, and on exchange, which does not. For detuned qubits the
odd block no longer commutes with itself at different times, so the odd crossings
become avoided and may leave their own, narrower features. Correlated baths would
modify the picture as well, since a common bath decouples the singlet. These
cases are beyond the present work.

\section{Conclusions}
\label{sec:conclusions}

We have studied the steady-state entanglement of two identical qubits with an
anharmonically modulated transverse coupling, comparing a time-independent
dressed master equation with quasienergy-resolved Floquet master equations. 
Although the modulation raises the ground-state concurrence of the cycle-averaged Hamiltonian, the steady-state concurrence does not increase. Instead, it stays close to its undriven value and vanishes at the even-sector Floquet anticrossings.

Parity and exchange symmetry select these resonances. Dissipation connects the
even and odd sectors, the even sector anticrosses, and the odd sector, which is
exactly solvable, only crosses and leaves no mark. The deaths require the
anticrossing gap to exceed the damping rate.

The Floquet and static descriptions place the resonances identically but differ
in the entanglement between them, by up to $0.2$ for an Ohmic bath and $0.5$ for a
flat one. We trace this difference to the basis of the dissipator rather than to
the secular approximation, and to channels that the drive moves through zero
frequency, which make the Floquet steady state sensitive to the low-frequency
shape of the bath. Breaking parity adds inter-sector resonances, which connects
these results to recent work on a parity-broken version of the model.

These predictions can be tested with parametrically modulated superconducting
qubits. The gaps lie in the MHz range, well above typical damping rates, and the
resonances and their selection rule are visible in the emission. Tuning the
damping or filtering the environment would further probe the effects of the dissipator
basis and of the bath identified here.

More generally, when a drive engineers an interaction, the symmetries decide
which resonances appear in the steady state, and the way the dissipator resolves
the Floquet spectrum decides how strongly they appear. Static and secular
treatments must be checked against both.

\begin{figure*}[t]
\centering
\includegraphics[width=\textwidth]{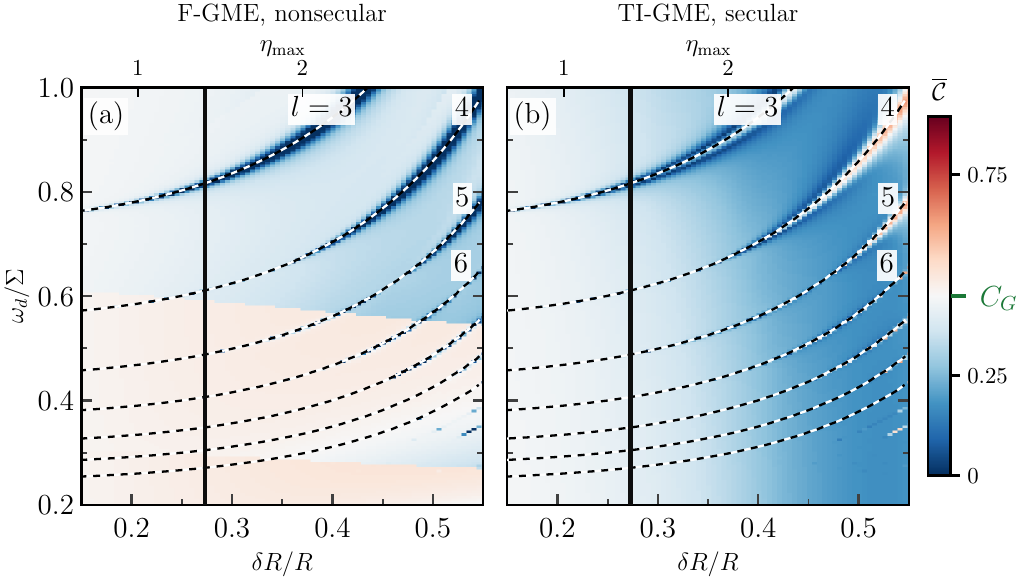}
\caption{As Fig.~\ref{fig:maps}, but for the flat bath, from (a)~the nonsecular
F-GME and (b)~the secular TI-GME, with the same $\gamma=10^{-3}\Sigma$ and grid. The
sharp edges in (a), near $\omega_d\simeq0.6\Sigma$ and $0.28\Sigma$, occur where an
inter-sector channel passes through zero frequency, where the flat spectral
density has a step. At $\omega_d=0.6\Sigma$ the maps reproduce the flat-bath curves
of Fig.~\ref{fig:main}.}
\label{fig:maps_flat}
\end{figure*}

\begin{acknowledgments}
This work was supported by the Natural Sciences and Engineering Research Council of
Canada (NSERC), the Canadian Foundation for Innovation (CFI), and Queen's
University, Canada. The authors acknowledge the use of  Claude
by Anthropic for assistance with manuscript preparation and coding. All
AI-assisted output was reviewed, corrected, and  verified by the authors.
\end{acknowledgments}

\vspace{1cm}

\appendix

\section{Flat-bath maps}
\label{app:flat}

Figure~\ref{fig:maps_flat} shows the maps of Fig.~\ref{fig:maps} for a flat bath.
The resonance arcs are the same, but sharp edges now cross the maps where an
inter-sector channel changes the sign of its frequency and its flat-bath rate
switches on or off. These edges are absent for the Ohmic bath, whose rate
vanishes continuously at zero frequency.

\section{Positivity}
\label{app:positivity}

\begin{figure*}[t]
\centering
\includegraphics[width=\textwidth]{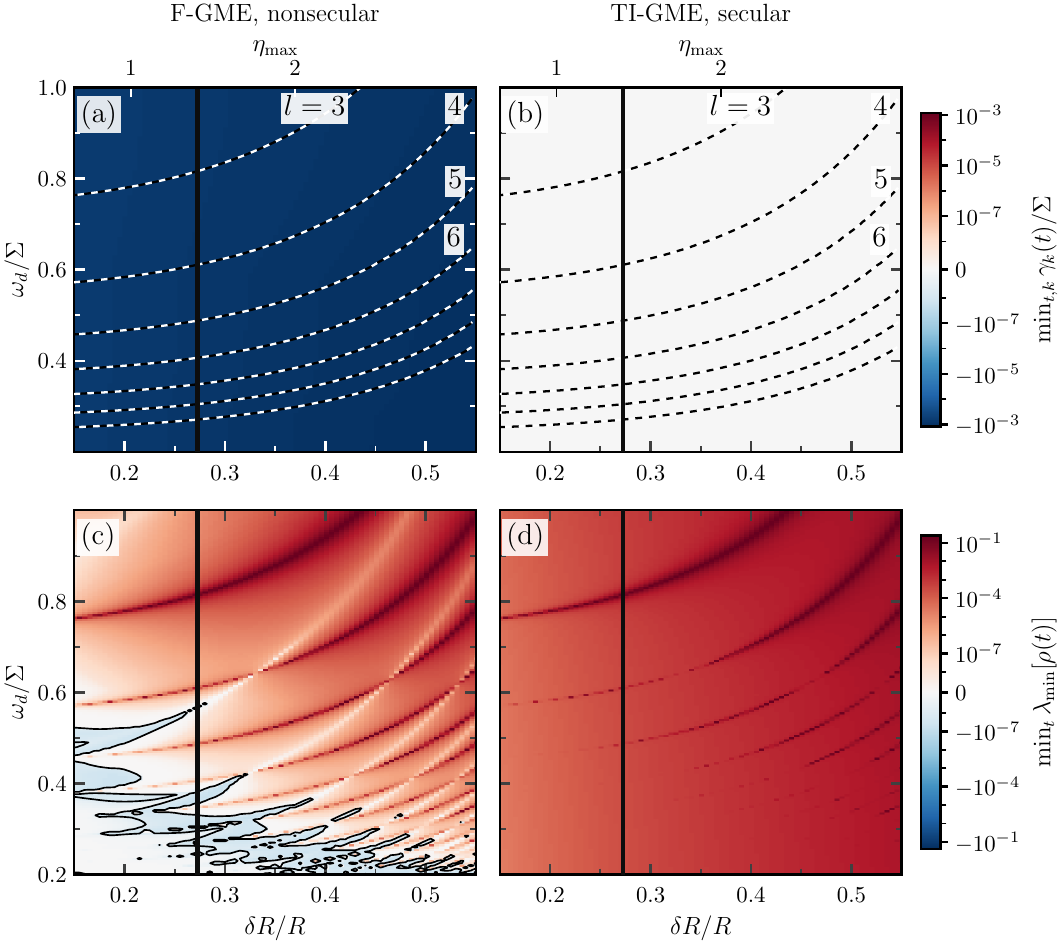}
\caption{Positivity diagnostics for the Ohmic maps of Fig.~\ref{fig:maps}.
(a),~(b)~smallest canonical rate of the generator, minimized over the cycle, and
(c),~(d)~smallest eigenvalue of $\rho_{\rm ss}(t)$, minimized over the cycle, for
the nonsecular F-GME (left) and the secular TI-GME (right). Solid contours in (c)
mark $-10^{-8}$.}
\label{fig:positivity}
\end{figure*}

A generator is of GKLS form when all its canonical (Kossakowski) rates are
non-negative; a steady state is physical when all eigenvalues of
$\rho_{\rm ss}(t)$ are non-negative. Figure~\ref{fig:positivity} shows both
diagnostics for the Ohmic maps. The secular TI-GME has non-negative rates to
machine precision and positive states, $\lambda_{\min}\ge1.8\times10^{-5}$. The
nonsecular F-GME has a negative canonical rate everywhere, between
$-1.2\times10^{-3}\Sigma$ and $-0.6\times10^{-3}\Sigma$, but its states are
physical to within $1.3\times10^{-7}$. The smallest eigenvalue is negative at
$15\%$ of the points, always with $|\lambda_{\min}|\le1.3\times10^{-7}$. Small negative eigenvalues
occur where the state is close to pure, not near the anticrossings where the
concurrence changes. The flat-bath maps of Fig.~\ref{fig:maps_flat} give the same
picture. The secular TI-GME again has non-negative rates and positive states,
$\lambda_{\min}\ge1.8\times10^{-5}$, and the nonsecular F-GME has negative canonical
rates between $-0.74\gamma$ and $-0.42\gamma$ but states physical to within
$1.5\times10^{-6}$.

\section{Comparison with Ref.~\cite{Pacino_Foquet-Resolved_2026}}
\label{app:pacino}

\begin{figure*}[t]
\centering
\includegraphics[width=\textwidth]{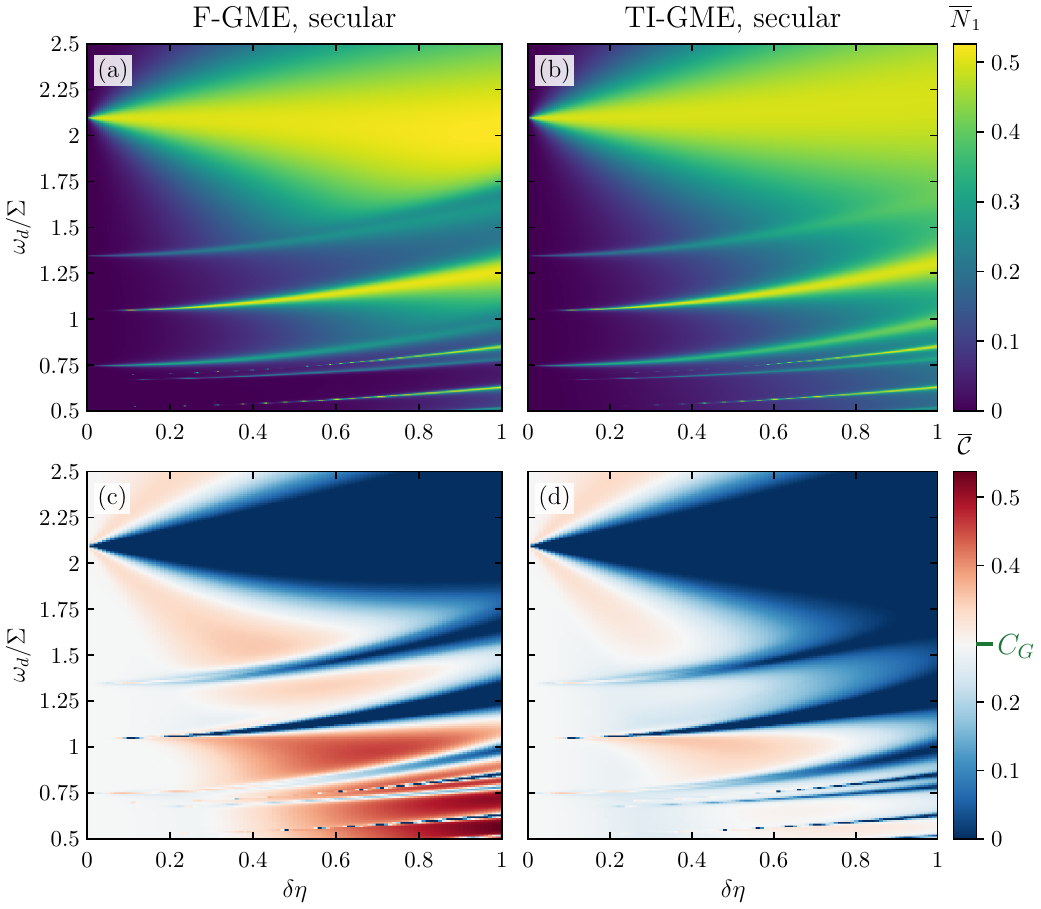}
\caption{Results for the model of Ref.~\cite{Pacino_Foquet-Resolved_2026}, with
$\eta=0.3$, $\epsilon_1=0.1\Sigma$, $\epsilon_2=0$, an Ohmic bath, $T=0$ and
$\gamma=10^{-3}\Sigma$. (a),~(b)~Cycle-averaged population $\Nbar_1$ of qubit~1 and
(c),~(d)~cycle-averaged concurrence $\Cbar$, versus $(\delta\eta,\omega_d)$, from the
secular F-GME (left) and the secular TI-GME (right). The axes are exchanged with
respect to Ref.~\cite{Pacino_Foquet-Resolved_2026}, and $\delta\eta$ corresponds to
their $A/\Delta$, the modulation amplitude in units of the qubit frequency
(Sec.~\ref{sec:harmonic}).}
\label{fig:pacino}
\end{figure*}

During the completion of this work we became aware of
Ref.~\cite{Pacino_Foquet-Resolved_2026}, which reports overlapping results for a
harmonically modulated coupling with a parity-breaking transverse field. Here we
show that we recover their results when we adopt the same model.
Figure~\ref{fig:pacino} shows our results for their model and parameters. Their Floquet--Born--Markov equation is
secularized between distinct Floquet channels and corresponds to our secular
F-GME. Their partial harmonic decomposition approach builds the dissipator from
the eigenstates of the static Hamiltonian, grouped by transition frequency, and
corresponds to our secular TI-GME. Both use the same Ohmic rate convention as
ours.

The populations in Figs.~\ref{fig:pacino}(a) and~\ref{fig:pacino}(b) reproduce
the period-averaged dressed populations of qubit~1 in Figs.~2(d) and~2(a) of
Ref.~\cite{Pacino_Foquet-Resolved_2026}, with the axes exchanged. For the
entanglement, Ref.~\cite{Pacino_Foquet-Resolved_2026} reports the concurrence at
fixed phases of the cycle, $t_0=0$ and $\tau/2$ (their Fig.~4), whereas we report
its cycle average. The two are not directly comparable point by point, but they
lead to the same conclusion. Where the two methods differ, the TI-GME gives less
entanglement than the F-GME at $92\%$ of the points of the map, consistent with
their finding that the static treatment underestimates the entanglement.

\section{Fourier coefficients and the odd sector}
\label{app:fourier}

The Fourier coefficients $g_m=g\,I_m$ of Eq.~\eqref{eq:gt},
$I_m=T^{-1}\int_0^T e^{\ii m\omega_d t}[1+r\sin(\omega_d t)]^{-3}\dd t$, follow
from contour integration on the unit circle, where only one root of
$1+r\sin\theta$ lies inside. With $s=\sqrt{1-r^2}$ and $q=r/(1+s)$,
\begin{equation}
I_m=(-\ii)^m\,q^{|m|}\,
\frac{(m^2-1)s^2+3|m|s+3}{2s^5},
\label{eq:Im}
\end{equation}
with $I_{-m}=I_m^*$. In particular $I_0=(2+r^2)/(2s^5)$,
$I_{\pm1}=\mp3\ii r/(2s^5)$ and
$I_{\pm2}=-3r^2/(2s^5)$. The coefficients decay as $q^{|m|}$, with
$q\to1$ as $r\to1$.

For $\omega_1=\omega_2$ the odd block of Eq.~\eqref{eq:H} is
$H(t)|_{\rm odd}=g(t)\,\mathcal{E}|_{\rm odd}$, with $\mathcal{E}\ket{U}=\ket{U}$
and $\mathcal{E}\ket{L}=-\ket{L}$. Since it commutes with itself at different
times, the Schr\"odinger equation integrates directly, giving
\begin{equation}
\ket{\psi_\pm(t)}=e^{\mp\ii g_0t}\,e^{\mp\ii\varphi(t)}\ket{U/L},\quad
\varphi(t)=\int_0^t[g(t')-g_0]\,\dd t',
\end{equation}
where $\varphi(t)$ is periodic. The Floquet states are $e^{\mp\ii\varphi(t)}\ket{U/L}$,
with quasienergies $\pm g_0$ modulo $\omega_d$. The odd branches are therefore the
static levels at the renormalized coupling, and they cross at
Eq.~\eqref{eq:oddloci}. The even block,
$H(t)|_{\rm even}=\tfrac{\Sigma}{2}(\sigma_z^{(1)}+\sigma_z^{(2)})+g(t)\sigma_x^{(1)}\sigma_x^{(2)}$,
satisfies $[H(t),H(t')]\propto[g(t')-g(t)]\,\sigma_y^{(1)}\sigma_y^{(2)}\neq0$ and has
no such solution. The numerical Floquet spectrum reproduces the odd
quasienergies to machine precision, and the exchange label is conserved to the
same accuracy.

\clearpage
\bibliography{main}

\end{document}